\documentclass[aps,prd,twocolumn,citeautoscript,reprint,preprintnumbers,superscriptaddress,nofootinbib]{revtex4}

\usepackage{amsmath,amssymb,bm,color,comment,tabularx,graphicx,ulem}
\usepackage[hidelinks=true]{hyperref}
\hypersetup{
  colorlinks = true,
  linkcolor  = blue,
  citecolor  = blue,
  urlcolor   = blue
}

\def\l{\ell}

\def\hatn{\Omega}
\def\hC{\widehat{C}}

\def\DS{\alpha}
\def\bn{\bm{\nabla}}
\def\ave#1{\langle #1 \rangle}
\def\vs #1{\vspace{#1\baselineskip}}
\newcommand\E{\hspace{0.1em}{\rm e}}
\newcommand{\bR}[1]{{\bm {\mathrm{#1}}} }
\newcommand{\D}[2]{\dfrac{d #1}{d #2}}
\newcommand{\mC}[1]{\mathcal{#1}}   
\newcommand{\dd}[2]{\,{\rm d}^{#1} #2 \,} 
\newcommand{\al}[1]{\ifdefined\compile \begin{align} #1 \end{align}\fi}
\newcommand{\INT}[5]{{\int_{#4}^{#5} \!\! \dd{#1}{#2} \,}}
\newcommand{\bc}{\begin{center}}
\newcommand{\ec}{\end{center}}

\newcommand{\compile}{}

\begin{document}

\ifdefined\compile 

\title{Detecting Black-Hole Binary Clustering via the Second-Generation Gravitational-Wave Detectors}

\author{Toshiya Namikawa}
\affiliation{Department of Physics, Stanford University, Stanford, California 94305, USA}
\affiliation{Kavli Institute for Particle Astrophysics and Cosmology, SLAC National Accelerator
Laboratory, Menlo Park, California 94025, USA}
\author{Atsushi Nishizawa}
\affiliation{Department of Physics and Astronomy, The University of Mississippi, University, Mississippi 38677, USA}
\affiliation{Theoretical Astrophysics 350-17, California Institute of Technology, Pasadena, California 91125, USA}
\author{Atsushi Taruya}
\affiliation{Center for Gravitational Physics, Yukawa Institute for Theoretical Physics, Kyoto University, Kyoto 606-8502, Japan}
\affiliation{Kavli Institute for the Physics and Mathematics of the Universe,
Todai Institutes for Advanced Study, the University of Tokyo, Kashiwa, Chiba 277-8583, Japan}

\date{\today}


\begin{abstract}
The first discovery of the gravitational-wave (GW) event, GW150914, suggests a higher merger rate of 
black-hole (BH) binaries. If this is true, a number of BH binaries will be observed via the 
second-generation GW detectors, and the statistical properties of the observed BH binaries can be 
scrutinized. A naive but important question to ask is whether the spatial distribution of BH binaries 
faithfully traces the matter inhomogeneities in the Universe or not. Although the BH binaries are 
thought to be formed inside the galaxies in most of the scenarios, there is no observational evidence 
to confirm such a hypothesis. Here, we estimate how well the second-generation GW detectors can 
statistically confirm the BH binaries to be a tracer of the large-scale structure by looking at the auto- and 
cross-correlation of BH binaries with photometric galaxies and weak-lensing measurements, finding that, 
with a 3 year observation, the $>3\sigma$ detection of a non-zero signal is possible if the BH merger 
rate today is $\dot{n}_0\agt100$ Gpc$^{-3}$yr$^{-1}$ and the clustering bias of BH binaries is 
$b_{\rm BH,0}\agt1.5$. 
\end{abstract}


\preprint{YITP-16-41}

\maketitle

\fi 


\section{Introduction}

The first discovery of the gravitational-wave (GW) event, GW150914, by aLIGO \cite{GW150914} opens 
a new window to astronomy and physics. The detected signal is consistent with GW emission from the 
coalescence of a black-hole (BH) binary at $z\simeq0.09$, demonstrating that the advanced detector has 
a sufficient sensitivity enough to detect GWs out to the distant universe. In the coming years, 
aVIRGO and KAGRA will join the network of the second-generation GW detectors \cite{Evans2014GRG} 
and will detect a large number of GW sources. In addition, the future ground- and space-based GW 
experiments such as the Einstein Telescope (ET) \cite{Punturo:2010}, 40 km LIGO \cite{Dwyer2015PRD}, 
eLISA \cite{Seoane:2013qna}, and DECIGO \cite{Sato:2009} are planning to greatly improve 
the sensitivities and realize cosmology with a large number of GW events at very high redshifts ($z>1$). 

From the cosmological point of view, one important aspect of the ongoing and future GW observations 
is that, using binary GW sources as {\it the standard sirens}, we will be able to measure 
the luminosity distance to each source with unprecedented precision 
\cite{Schutz1986,Petiteau2011ApJ,Cutler:2009qv,Nishizawa:2011eq,Sathyaprakash:2009xt,Taylor:2012db,Hirata:2010,Shang2011MNRAS,Camera:2013xfa}.
In particular, we have recently shown in Ref.~\cite{Namikawa:2016} that, without electromagnetic followup 
observations (i.e., redshift information), these standard sirens can be used to probe 
the large-scale structure (LSS) of the Universe at very high redshift ($z\agt 2$) where 
the identification of the electromagnetic counterpart is challenging. It will provide a way to 
tightly constrain the primordial non-Gaussianity of the large-scale matter fluctuations and to 
directly probe the matter inhomogeneities by cross-correlating with weak-lensing signals. 
Further, assuming that the binary GW sources are a good tracer of LSS, Ref.~\cite{Oguri:2016} 
explored the feasibility to cross-correlate the GW sources with spectroscopic galaxies and showed 
that the distance-redshift relation for GW sources can be estimated accurately without the followup 
observation of each GW source. 

While the methods proposed in Refs.~\cite{Namikawa:2016,Oguri:2016} are quite promising, 
the validity of the assumption that the binary GW sources fairly trace the matter inhomogeneities 
is largely unknown, because there is so far no observation to confirm the clustering hypothesis. 
If the GW sources are the primordial BH dark matter \cite{Clesse:2015,Bird:2016,Clesse:2016}, 
the clustering of the GW sources would be different from that of the astrophysical BH binaries 
(see, e.g., Refs.~\cite{Bekczynski:2010,Kinugawa:2014} for quantitative predictions). Furthermore, 
even with future electromagnetic observations, it would be rather difficult to identify robustly 
the electromagnetic counterparts, from which we can know what kind of galaxies or components 
(i.e., dark matter or baryon) BH binaries are likely to trace. These issues should be addressed by 
statistically measuring the clustering signal of GW sources themselves \cite{Namikawa:2016} and/or 
by cross-correlating with other independent mass tracers such as galaxy clustering and 
weak gravitational lensing \cite{Namikawa:2016,Oguri:2016}. It is therefore important to test or 
verify the clustering hypothesis of GW sources from the ongoing/upcoming 
GW observations prior to the future cosmological studies with third-generation GW detectors. 

In this paper, extending the analysis in Ref.~\cite{Namikawa:2016}, we shall discuss the 
feasibility to detect the clustering signal of binary GWs via a network of the second-generation 
GW detectors. In particular, we will focus on the BH binaries similar to GW150914. Indeed, the 
first GW detection enlarges the future prospect for measuring GWs from BH binaries and suggests 
a rather higher merger rate, $2$--$400$ Gpc$^{-3}$ yr$^{-1}$, indicating that even the 
second-generation GW detectors have a potential to detect the clustering of BH binary sources. 

This paper is organized as follows. In Sec.~\ref{Sec.2}, we begin by reviewing the statistical 
observables of the clustering signal, namely, the angular power spectrum, which are estimated both 
from auto- and cross-correlation of the BH binary clustering with clustering and weak-lensing 
signals from galaxy photometric surveys and cosmic microwave background (CMB) measurements. We 
then describe our assumptions on the noise properties of each observable in Sec.~\ref{Sec.3}. The 
significance of detecting the BH binary clustering is estimated in Sec.~\ref{Sec.4}. Finally, 
Sec.~\ref{Sec.5} is devoted to a summary and discussion.

Throughout the paper, the power spectra of the matter fluctuations are computed with the CMB Boltzmann 
code {\tt CAMB} \cite{CAMB}, assuming the flat Lambda-cold dark matter (CDM) model with fiducial 
cosmological parameters consistent with the 7 year WMAP results \cite{Komatsu:2010fb}. We use 
{\tt Halofit} for computing the nonlinear matter power spectrum \cite{Smith:2003,Takahashi:2012}.
We adopt the natural unit. 

\section{Observables} \label{Sec.2}

To statistically detect the clustering signals from BH binaries, we consider the angular power 
spectra between observables obtained from GW detectors, galaxy imaging surveys, and CMB experiments. 
In a spatially flat cosmological model, the auto- and cross-angular power spectra are related to the 
three-dimensional power spectrum of the matter fluctuations through 
(see, e.g., Refs.~\cite{Hu:2001fb,Namikawa:2011,Yamauchi:2013})
\al{
	C^{\rm XY}_\l &= 4\pi \INT{}{\ln k}{}{0}{\infty} 
		\INT{}{\chi}{}{0}{\infty} j_\l(k\chi) \INT{}{\chi'}{}{0}{\infty} j_\l(k\chi') 
		\notag \\
	&\quad \times W^{\rm X}(k,\chi) W^{\rm Y}(k,\chi') \Delta_{\rm m}(k;\chi,\chi')
	\,,
}
with the quantity $\chi$ being the comoving radial distance. Here, X and Y denote the observables from 
either the BH binary clustering ($s$), galaxy clustering ($g$), galaxy weak lensing ($\gamma$), or 
weak lensing of CMB ($\phi$). The function $\Delta_{\rm m}(k;\chi,\chi')$ is the dimensionless power 
spectrum of the matter density fluctuations, and $j_\l$ is the spherical Bessel function. The function 
$W^{\rm X}(k,\chi)$ is the weight function of an observable X, the functional form of which will be 
specified below.

\subsection{Clustering of BH binaries}

BH binaries are the representative candidate of the GW standard sirens observed via the 
second-generation GW detectors. If the BH binaries trace the LSS, their spatial distribution would 
have a characteristic pattern, the statistical properties of which are related to those of the LSS. 
In principle, with the GW observation alone, one can map out the three-dimensional clustering of BH 
binaries; however, we do not use distance information in our analysis. This is because the observable 
redshift for BH binaries similar to GW150914 will be limited to $z\alt 0.3$ for the second-generation 
detectors \cite{GW150914rate}, and the expected number of GW events is thus not so large 
($\mC{O}(10^2)-\mC{O}(10^3)$). To enhance the detection significance, we therefore consider the 
two-dimensional map, i.e., the angular distribution of BH binaries projected onto the sky. 

Ignoring the lensing contribution to the luminosity distance, which is shown to be subdominant in the
two-dimensional sky map of the GW sources \cite{Namikawa:2016}, the weight function of BH binaries 
becomes 
\al{
	W^s(\chi) = \D{n_{\rm BH}}{\chi}(\chi)\, b_{\rm BH}(z(\chi)) \,, \label{eq:weight_BH}
}
where $dn_{\rm BH}/d\chi$ is the radial distribution of BH binaries given by 
\al{
	\D{n_{\rm BH}}{\chi}(\chi) = \frac{1}{{N_{\rm BH}}}\,\,T_{\rm obs}\,\dot{n}_0\frac{\chi^2}{1+z(\chi)} 
	\,, \label{Eq:nBH}
}
with $T_{\rm obs}$ and $\dot{n}_0$ being the observation time and the merger rate today, respectively. 
The quantity, $N_{\rm BH}$, is the total number of BH binaries per steradian so as to give 
$\int {\rm d}\chi\,(dn_{\rm BH}/d\chi) =1$. Here, we assume the constant merger rate, since the 
observable redshift of BH binaries via the second-generation detectors will be $z\alt 0.3$. In 
Eq.~(\ref{eq:weight_BH}), we introduce $b_{\rm BH}(z)$, which represents the clustering bias of BH 
binaries. For BH binaries associated with galaxies, the bias factor $b_{\rm BH}(z)$ simply reflects 
the galaxy bias, and it may vary with time. Below, assuming the functional form of 
$b_{\rm BH}(z)=b_{\rm BH,0}(1+z)^{1/2}$ \cite{Rassat:2008}, we estimate the detectability of the 
clustering signal, and discuss its sensitivity to $b_{\rm BH,0}$. Since the redshift range we consider 
in the analysis below is very narrow, the evolution of the clustering bias does not significantly 
alter our results.

\subsection{Clustering of galaxies}

As one of the independent LSS tracers, we consider the photometric galaxies to cross-correlate with 
BH binaries. Similar to the BH binary clustering, the weight function of the galaxy clustering 
becomes (e.g., Ref.~\cite{Namikawa:2011})  
\al{
	W^{\rm g}(\chi) = \D{n_{\rm gal}}{\chi}(z(\chi)) \,b_{\rm gal}(z(\chi)) \,, 
}
where $dn_{\rm gal}/d\chi$ and $b_{\rm gal}(z)$ are the normalized number density and bias factor of 
the galaxies, respectively. For simplicity, we assume the same bias evolution as the BH binary case; 
$b_{\rm gal}(z)=b_{\rm gal,0}(1+z)^{1/2}$ \cite{Rassat:2008}. For the normalized distribution 
function, we adopt the form \cite{Euclid}
\al{
	\D{n_{\rm gal}}{\chi}(z) = \frac{3z^2}{2z_0^3}\exp\left[-\left(\frac{z}{z_0}\right)^{3/2}\right] H(z)
	\,, \label{Eq:ngal}
}
where the parameter $z_0$ is related to the mean redshift $z_{\rm m}$ through $z_{\rm m}=1.412z_0$ 
\cite{Euclid} and $H(z)$ is the expansion rate. The last factor $H(z)$ simply comes from the 
conversion between $z$ and $\chi$.

\subsection{Weak lensing of galaxies}

The weak lensing of galaxies also provides a way to probe LSS, and the measurement of this can be used 
to cross-correlate with BH binaries. The key observable of the weak lensing considered here is the 
shear $\gamma(\hatn)$, which is obtained by measuring ellipticities of each galaxy image. The shear is 
related to the gravitational potential of the matter density fluctuations, and the weight function is 
thus expressed as (e.g., Refs.~\cite{Bartelmann:2001,Yamauchi:2013})
\al{
	W^\gamma(k,\chi) &= \sqrt{\frac{(\l+2)!}{(\l-2)!}}\frac{3\Omega_{\rm m}H_0^2(1+z)}{2k^2}
		\notag \\
	&\times \INT{}{\chi'}{}{\chi}{\infty} \frac{\chi'-\chi}{\chi'\chi} 
		\D{n_{\rm gal}}{\chi'}(z(\chi')) \,, 
}
where $H_0$ and $\Omega_{\rm m}$ are the present Hubble parameter and the density parameter of 
the mass, respectively. For the distribution of source galaxies $dn_{\rm gal}/d\chi$, 
we adopt the same functional form as given in Eq.~(\ref{Eq:ngal}), since the galaxies identified with 
photometric surveys are also used for the weak-lensing measurement.

\subsection{Weak lensing of CMB}

The gravitational lensing induced by the LSS also affects the CMB at each angular position. With the 
help of the reconstruction technique, we can probe the LSS from the distortion of the primary CMB 
anisotropies. The lensing effect on CMB anisotropies is expressed as a remapping by the two-dimensional 
deflection vector $\bm{d}=\bn\phi$, where $\phi$ is so-called the CMB lensing potential 
(e.g., Ref.~\cite{Lewis:2006fu}). This lensing potential is an observable reconstructed from a CMB map 
by utilizing the characteristic mode coupling between lensed CMB anisotropies 
(e.g., Ref.~\cite{Hu:2001kj}). Since the lensing comes from the last scattering surface of CMB photon 
which is approximately described by the single-source plane, the weight function of the lensing 
potential $\phi$ is given by (e.g., Refs.~\cite{Lewis:2006fu,Yamauchi:2013})
\al{
	W^{\phi}(k,\chi) = \frac{3\Omega_{\rm m}H_0^2(1+z)}{2k^2}\frac{\chi_*-\chi}{\chi_*\chi} \qquad (\chi\leq\chi_*) \,, 
}
and becomes zero otherwise. Here, the quantity $\chi_*$ indicates the comoving radial distance to the 
last-scattering surface.

\section{Detection significance of clustering signal} \label{Sec.3}

In the absence of observational evidence for BH binaries to be a good tracer of the matter inhomogeneities, 
we test the null hypothesis that the distribution of BH binaries is spatially homogeneous. 
We investigate the significance of rejecting this null hypothesis 
(hereafter, we call it detection significance shortly, following the convention, e.g., Ref.~\cite{Chisari:2014}).

In the case, using the GW data alone, the statistical significance to reject null hypothesis is 
quantified by (see, e.g., Refs.~\cite{PB14:phi,Chisari:2014})
\al{
	\DS^2_{ss} = \sum_\l\frac{2\l+1}{2} \left[\frac{C^{ss}_\l}{N_\l^{ss}}\right]^2 \, \label{Eq:auto}
}
with $N_\l^{ss}$ being the noise spectra for the clustering signal of BH binaries given later. We here 
assume a full-sky GW observation. Note that Eq.~(\ref{Eq:auto}) slightly differs from the usual 
definition of the signal-to-noise ratio, since we consider the null hypothesis for the clustering of 
BH binaries. On the other hand, if one uses other cosmological probes ($X=g$, $\gamma$, or $\phi$) 
to cross-correlate with GW data ($s$), the statistical significance to reject the null hypothesis is 
estimated from
\al{
	\DS^2_{sX} = f^{sX}_{\rm sky} \sum_\l(2\l+1)\frac{(C^{sX}_\l)^2}{(C^{XX}_\l+N^{XX}_\l)N_\l^{ss}}
	\,, \label{Eq:cross}
}
where $f^{sX}_{\rm sky}$ denotes the fractional sky coverage of the other cosmological probes. 
$N_\l^{XX}$ is the noise power spectrum of each observable.  

Combining all the observables including GW observations, photometric galaxies and weak lensing of 
galaxies and CMB, the total detection significance is written as 
\al{
	\DS^2_{\rm tot} = \sum_\l(2\l+1)\sum_{i,j}\min(f^i_{\rm sky},f^j_{\rm sky})\{\bR{C}_\l^{-1}\}_{i,j} 
	\,, \label{Eq:tot}
}
where the indices, $i$ and $j$, are either $ss$, $sg$, $s\gamma$ or $s\phi$. The covariance matrix 
$\bR{C}_\l$ is defined as $\{\bR{C}_\l\}_{i,j}=\ave{\hC^i_\l\hC^j_\l}_{\rm null}/C^i_\l C^j_\l$, where 
$\hC^i_\l$ is the measured power spectrum including noise and $\ave{\cdots}_{\rm null}$ is the 
ensemble average but ignores the cosmic variance from the BH binary clustering. The upper triangular 
elements of the covariance matrix are given by
\al{
	\bR{C}_\l = \begin{pmatrix} 
		\frac{(N^{ss}_\l)^2}{(C^{ss}_\l)^2} & 0 & 0 & 0 \\ 
		& \frac{(C^{gg}_\l+N^{gg}_\l)N_\l^{ss}}{(C^{sg}_\l)^2} & 
		\frac{C^{g\gamma}_\l N_\l^{ss}}{C_\l^{sg}C_\l^{s\gamma}} & 
		\frac{C^{g\phi}_\l N_\l^{ss}}{C_\l^{sg}C_\l^{s\phi}} \\ 
		& & \frac{(C^{\gamma\gamma}_\l+N^{\gamma\gamma}_\l)N_\l^{ss}}{(C^{s\gamma}_\l)^2} & 
		\frac{C^{\gamma\phi}_\l N_\l^{ss}}{C_\l^{s\gamma}C_\l^{s\phi}} \\ 
		& & & \frac{(C^{\phi\phi}_\l+N^{\phi\phi}_\l)N_\l^{ss}}{(C^{s\phi}_\l)^2} 
	\end{pmatrix} \,.
}

Let us describe noise properties of each observable. For the GW observations of BH binaries, the 
dominant noise contribution would be the shot noise in source counting due to a limited number of BH 
binaries (not the photon counting shot noise in the GW detector). Further, the limited sky 
localization of each GW source restricts the sensitivity to the angular clustering. Thus, we consider 
the following noise spectrum for the BH binaries:
\al{
	N_\l^{ss}(z) = \frac{1}{N_{\rm BH}}\E^{\l(\l+1)\theta^2(z)/8\ln 2} \,.
}
The shot-noise contribution given above is convolved with the two-dimensional Gaussian window function 
with the full width of half maximum (FWHM) of $\theta$ \cite{Knox:1999}, which represents the angular 
resolution due to the limited sky localization of the GW sources. Note that the FWHM $\theta$ varies 
with the redshift of GW sources. Based on Fig.~5 of Ref.~\cite{Sidery2014PRD}, in which the angular 
resolution of each binary source is estimated assuming the second-generation detectors, we adopt the 
fitting form $\theta(z) \approx 45\,{\rm deg}/\rho_{\rm net}(z)$, where $\rho_{\rm net}$ is the 
detector-network signal-to-noise ratio (SNR) \footnote{This expression corresponds to 
$\l_{\rm max}(z)\approx 4.0 \times \rho_{\rm net}(z)$ in terms of the multipole.}. Note that, as 
discussed in Ref.~\cite{Namikawa:2016}, the uncertainties of the luminosity distance measurements 
are negligible compared to the above shot noise. This indicates that our results are insensitive to 
the noise spectrum of the BH binaries. On the other hand, a non-Gaussian localization errors modifies 
the functional form of the above shot-noise power spectrum especially at small angular scales, though 
the clustering signals at small scales do not so affect the resultant detection significance. To 
include a realistic non-Gaussian error, we need to characterize the location and orientation of each 
detector, and the impact of a realistic localization error remains our future work.

\begin{figure}[t]
\bc
\ifdefined\compile
\includegraphics[width=8.5cm,clip,height=6cm]{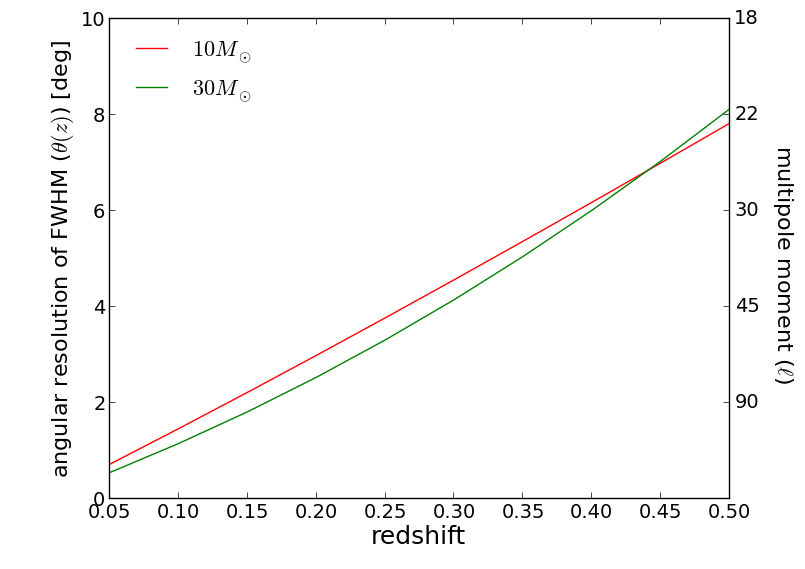}
\fi
\caption{
Angular resolution of GW sources, $\theta(z)$, achievable with a network of second-generation 
detectors. The results are plotted as function of source redshift, assuming BH binary of 
$10$-$10\,M_\odot$ (red) and $30$-$30\,M_\odot$ (green).
}
\label{Fig:theta}
\ec
\end{figure}

Using the restricted 1.5 Post Newtonian (PN) waveform of the BH binary and sensitivity curve for the
aLIGO detector, we estimate $\rho_{\rm net}$ as a function of redshift for the BH binary with $10$-
$10\,M_\odot$ and $30$-$30\,M_\odot$ \footnote{Although this paper shows the case with the inspiral 
component alone, we also compute the SNR including the inspiral-merger-ringdown waveform and find 
it leads to $20$\%--$30$\% enhancement of the SNR, improving the pointing of GW sources by $20$\%--
$30$\%. However, the effect of this improvements is negligible in our estimate of the detection 
significance for $30$-$30\,M_\odot$ systems at low z considered in this paper.}. The resultant angular 
resolution $\theta$ is shown in Fig.~\ref{Fig:theta}. Although the angular resolution to each GW 
source becomes degraded as the redshift increases, this degradation is quantitatively insensitive to 
the choice of the fiducial BH masses. A careful reader may wonder why these two curves cross at 
$z=0.45$. This is due to a redshift effect, which is significant for massive binaries. We checked that 
our estimate of the detection significance is robust against the choice of the fiducial BH binary 
mass. Hence, we will present below the results with BH binaries of $30$-$30\,M_\odot$.  

As for the other LSS probes to be cross-correlated with BH binaries, the shot-noise contribution is 
the main noise source of the photometric galaxy measurements apart from the cosmic variance. Thus, 
similar to the BH binary case, we have 
\al{
	N_\l^{gg} = \frac{1}{N_{\rm gal}} \,, 
}
Here, $N_{\rm gal}$ is the number of galaxies per steradian. On the other hand, the main noise source 
in the weak-lensing measurement of galaxies is the intrinsic scatter of each galaxy image 
(i.e., shape noise), and the noise power spectrum becomes 
\al{
	N_\l^{\gamma\gamma} = \frac{\sigma_{\gamma}^2}{N_{\rm gal}} \,, 
}
where $\sigma_{\gamma}$ is the intrinsic rms shear. We adopt $\sigma_\gamma=0.2$ for later analysis
\cite{DES,Euclid}. Finally, for the weak lensing of CMB, the dominant noise contribution (called the 
reconstruction noise, denoted by $N_\l^{\phi\phi}$) is computed using the formula given in 
Ref.~\cite{Smith:2010gu}, which is based on the maximum-likelihood lensing reconstruction 
\cite{Hirata:2003ka}.

\section{Testing clustering hypothesis of BH binaries} \label{Sec.4}

\subsection{Setup}

\begin{table}
\bc
\caption{
The galaxy survey specification: the total number of galaxies per square 
arcminute $N_{\rm gal}$, mean redshift $z_{\rm m}$, and fraction of the sky coverage $f_{\rm sky}$. 
The values for DES and Euclid are taken from Refs.~\cite{DES} and \cite{Euclid}, respectively, 
while we denote Pan-STARRS as a wide shallow galaxy survey \cite{Cai:2009}. 
}
\label{Table:gal} \vs{0.5}
\begin{tabular}{lccc} \hline 
 & $N_{\rm gal}$ [arcmin$^{-2}$] & $z_{\rm m}$ & $f_{\rm sky}$ \\ \hline 
DES        & 12 & 0.68 & 0.125 \\ 
Euclid     & 30 & 0.90 & 0.500 \\ 
Pan-STARRS & 1  & 0.50 & 0.750 \\ \hline
\end{tabular}
\ec
\end{table}

To quantitatively estimate the statistical significance, we shall specify several parameters for each 
observable given in the previous section. 

First of all, we consider a network of three second-generation GW detectors with design sensitivity 
given in Ref.~\cite{Sathyaprakash:2009} and 3 year observation ($T_{\rm obs}=3$yr). The merger rate of 
BH binaries at present, $\dot{n}_0$, is one of the key parameters, but it has still large uncertainty, 
$\dot{n}_0=2-400$\, Gpc$^{-3}$yr$^{-1}$. The clustering bias parameter, $b_{\rm BH,0}$, which 
indicates how significantly the clustering of the GW sources trace the matter inhomogeneities is also 
unknown. While we choose $\dot{n}_0=100\,$Mpc$^{-3}$yr$^{-1}$ and $b_{\rm BH,0}=1.5$ as a canonical 
setup and estimate the combined detection significance $\DS_{\rm tot}$, the detection significances 
for each single measurement, $\DS_{ss}$ and $\DS_{sX}$, are found to simply scale as 
\al{
	\DS_{ss} &= \DS^0_{ss}\left(\frac{b_{\rm BH,0}}{1.5}\right)^2
		\left(\frac{T_{\rm obs}\dot{n}_0}{3\times 100 {\rm \,Gpc^{-3}}}\right) \,, \label{Eq:auto-dep}\\
	\DS_{sX} &= \DS^0_{sX}\left(\frac{b_{\rm BH,0}}{1.5}\right)
		\left(\frac{T_{\rm obs}\dot{n}_0}{3\times 100 {\rm \,Gpc^{-3}}}\right)^{1/2} \,. \label{Eq:cross-dep}
}
Hence, we will present the estimated results of $\DS^0_{ss}$ and $\DS^0_{sX}$ for the 
single-measurement cases. Note that for the canonical setup the total number of BH binaries detected 
by the full-sky observation is estimated to be $N_{\rm BH}=549$ $(1,617)$ at $z\leq 0.2$ $(0.3)$. As 
we will see below, with such a small number of events, the GW data alone (i.e., autocorrelation of BH 
binaries) cannot give a statistically significant detection, and the cross-correlation with other LSS 
data is indispensable. In such a case, the total detection significance, $\DS_{\rm tot}$, is mostly 
determined by the cross-correlation, and thus $\DS_{\rm tot}$ approximately follows the same scaling 
law as shown in Eq.~(\ref{Eq:cross-dep}). 

As other independent LSS probes, we consider three representative surveys for the clustering and 
weak lensing of photometric galaxies; the Dark Energy Survey (DES) \cite{DES}, Euclid \cite{Euclid}, 
and the Panoramic Survey Telescope and Rapid Response System (Pan-STARRS) $3\pi$ survey
\footnote{http://pan-starrs.ifa.hawaii.edu/public/home.html}. The parameters needed to compute the 
signal and noise spectra are summarized in Table \ref{Table:gal}. Note that in all three cases we 
assume $b_{\rm gal,0}=1.0$ and adopt $\sigma_{\gamma}=0.2$, but our results are insensitive to the 
choice of these, as we will discuss later. 

Finally, for the weak lensing of CMB, a relevant experiment at the time of the second-generation GW 
detectors would be the CMB Stage-III experiment such as Advanced ACT \cite{AdvACT} and Simons Array 
\cite{SA}, which will achieve a nearly half-sky observation ($f_{\rm sky}=0.5$). We assume a $5\mu$K 
arcmin white noise with a beam size of $1$ arcmin. To precisely reconstruct the lensing potential 
involved in the small-scale CMB anisotropies, the multipoles up to $\l=3,000$ are used. The 
reconstruction noise $N_\l^{\phi\phi}$ is then computed based on this setup.

\subsection{Results}

\begin{table}[tb]
\bc
\caption{
Detection significance of the clustering of BH binaries with the 3 year observation from the 
second-generation detector network, using GW data alone ($ss$), the cross-correlation with the galaxy 
clustering ($gs$), the galaxy weak lensing ($\gamma s$), and the CMB weak lensing ($\phi s$). To be 
precise, numerical values presented below represent the significance to reject the null hypothesis 
that BH binaries are homogeneously distributed, based on Eqs.~\eqref{Eq:auto}-\eqref{Eq:tot}. The 
upper part of the table shows the coefficient of the auto- and cross-correlation measurement, 
$\DS^0_{ss}$ and $\DS^0_{sX}$, defined in Eqs.~\eqref{Eq:auto-dep} and \eqref{Eq:cross-dep}, 
respectively. The last row lists the total detection significance, $\DS_{\rm tot}$. In all cases, the 
present-day merger rate of BH binaries is set to $100$ Gpc$^{-3}$yr$^{-1}$, and we assume the 
clustering bias of $b_{\rm BH,0}=1.5$ with the maximum redshift of BH binaries, $z_{\rm max}=0.2$. For 
comparison, the parentheses show the results in the most optimistic case ($\dot{n}_0=400$ Gpc$^{-3}$yr$^{-1}$). 
}
\label{Table:SNR} \vs{0.5}
\begin{tabular}{l|c|c|c} \hline 
           & \multicolumn{3}{c}{Detection Significance} \\ \hline \hline
$ss$       & \multicolumn{3}{c}{0.495 (1.98)} \\ \hline 
$\phi s$   & \multicolumn{3}{c}{0.972 (1.94)} \\ \hline 
 & $\times$ DES & $\times$ Euclid & $\times$ Pan-STARRS \\ \hline
$gs$       & 1.77  (3.55) & 3.58 (7.16) & 4.47 (8.93) \\ \hline
$\gamma s$ & 0.971 (1.94) & 1.93 (2.87) & 1.44 (2.87) \\ \hline \hline
 & \multicolumn{3}{c}{Total} \\ \hline
 & 1.85 (4.07) & 3.63 (7.46) & 4.50 (9.16) \\ \hline \hline
\end{tabular}
\ec
\end{table}

\begin{figure}[t]
\bc
\ifdefined\compile
\includegraphics[width=8.5cm,clip,height=6cm]{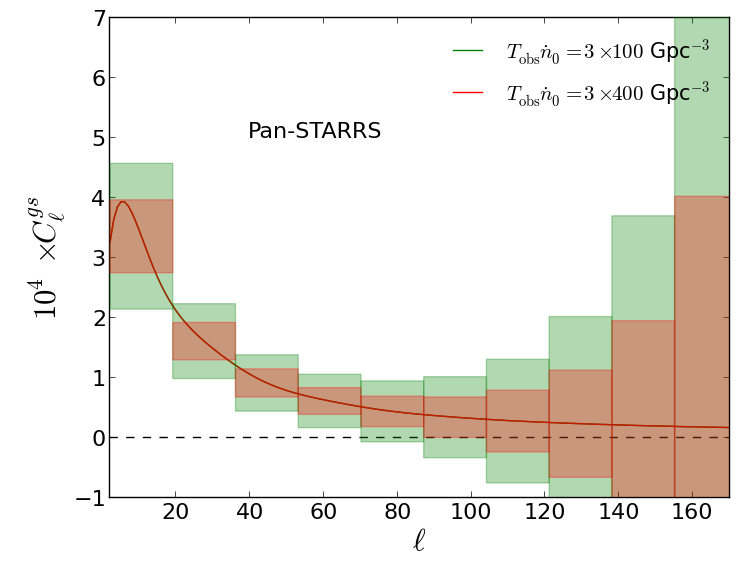}
\fi
\caption{
The cross angular power spectrum between the clustering of BH binaries and galaxies. The error boxes 
are computed assuming Pan-STARSS and with $T_{\rm obs}\dot{n}_0=3\times100$Gpc$^{-3}$ (green) or 
$3\times 400$Gpc$^{-3}$ (red). The maximum redshift of the GW sources is $z_{\rm max}=0.2$. 
}
\label{Fig:clnl}
\ec
\end{figure}

\begin{figure}[t]
\bc
\ifdefined\compile
\includegraphics[width=8.5cm,clip,height=6cm]{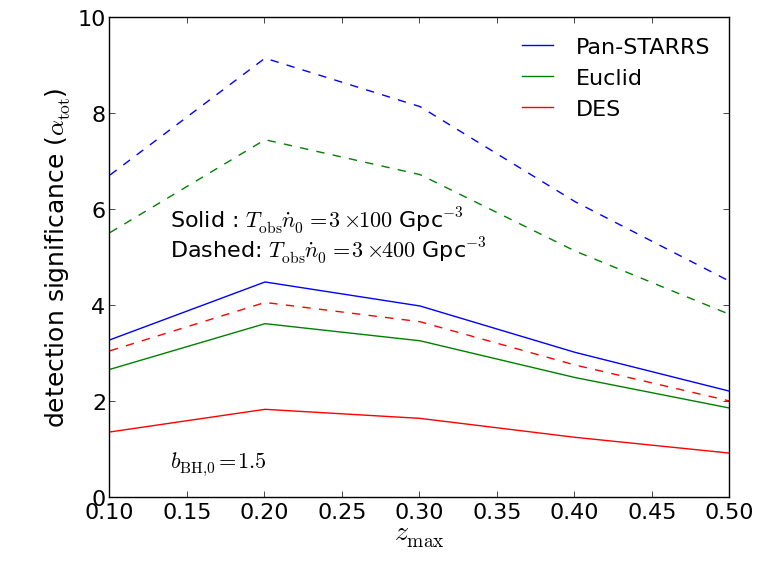}
\fi
\caption{
The combined result of the detection significance, $\DS_{\rm tot}$, as a function of the maximum 
redshift of BH binaries, $z_{\rm max}$, assuming the 3 year observation of the second-generation 
detector network. The quantity $\DS_{\rm tot}$ precisely implies the significance of rejecting 
the hypothesis that the distribution of BH binaries is spatially homogeneous. Blue, green, and red 
lines, respectively, show the result combined with Pan-STARRS, Euclid, and DES, for 
$T_{\rm obs}\dot{n}_0=3\times100$Gpc$^{-3}$ (solid) and $3\times400$Gpc$^{-3}$ (dashed).
}
\label{Fig:zmax}
\ec
\end{figure}

Let us first show the estimated values of the detection significance for each measurement. Table 
\ref{Table:SNR} summarizes the auto- ($\DS_{ss}^0$) and cross-correlation ($\DS_{sX}^0$) results in 
the canonical setup. Here, to compute the detection significance, the sample of BH binaries is 
restricted to $z\leq z_{\rm max}=0.2$. This is because samples at high redshifts are prone to have a 
poor angular resolution (see Fig.~\ref{Fig:theta}). Figure \ref{Fig:clnl} shows the angular power 
spectrum between the clustering of BH binaries and galaxies in the case of Pan-STARRS, with expected 
errors at each multipole bin. The results suggest that the clustering signal is very hard to detect 
by the GW data alone, but cross-correlating with other LSS probe enlarges the capability of detecting 
the clustering signal. In particular, the cross-correlation with photometric galaxies can give a 
higher detection significance, and a shallow but wide-field survey like Pan-STARRS will be able to 
give a solid detection. On the other hand, lensing measurements of both galaxies and CMB do not help 
so much to improve the detection significance. This is partly because the lensing signal is basically 
generated by the matter fluctuations at high redshifts. In this sense, the cross-correlation with 
the BH binary clustering at low redshifts is not optimal. Indeed, increasing the maximum redshift of 
the BH binary samples to $z_{\rm max}=0.3$, the detection significance is improved by $20$\%--$40\%$ 
from the lensing measurements ($20\%$ for DES, $21\%$ for Euclid, $43\%$ for Pan-STARRS, and $26\%$ 
for CMB Stage-III). 

In Table \ref{Table:SNR}, the combined results of both the auto- and cross-correlations for each 
survey are shown. Further, in Fig.~\ref{Fig:zmax}, the combined detection significance is plotted as 
a function of the maximum redshift of BH binaries, $z_{\rm max}$. In the most optimistic case with 
the merger rate of $\dot{n}_0=400\,$Gpc$^{-1}$yr$^{-1}$ (dashed lines), the detection of the 
clustering signal is fairly likely from the cross-correlation of BH binaries at $z\lesssim0.2-0.3$ 
with Pan-STARRS and Euclid. As is shown in Table \ref{Table:SNR}, the detection significance is mostly 
determined by cross-correlating with photometric galaxies. This implies that making full use of 
cross-correlations, we still have a chance to detect the clustering signal of BH binaries even with 
smaller merger rates, since the detection significance approximately scales as $\dot{n}_0^{1/2}$.  

Finally, note that the detection significances of the cross-correlation are basically limited by the 
shot noise of the GW sources and the cosmic variance of galaxies or lensing measurement. That is, 
Eq.~(\ref{Eq:cross-dep}) can be recast as
\al{
	\DS^2_{sX} \simeq f_{\rm sky} \sum_\l(2\l+1)\frac{(C^{sX}_\l)^2}{C^{XX}_\l N_\l^{ss}}
	\,. \label{Eq:cv}
}
This suggests that the results become nearly insensitive to the clustering bias of galaxies, 
$b_{\rm gal,0}$, and the intrinsic rms shear, $\sigma_{\gamma}$. Even though the galaxy number density 
is reduced to $0.1$ per square arcmin, Eq.~(\ref{Eq:cv}) would be valid for the cross-correlation with 
photometric galaxies, indicating that a wide-field survey is preferable to enhance the detection 
significance. In this respect, a cross-correlation with Large Synoptic Survey Telescope \cite{LSST} 
also helps to detect the BH binary clustering, and the detection significance will be rather 
comparable to that of Euclid. On the other hand, the cross-correlation with CMB lensing is still not 
useful to detect the BH binary clustering. This is true even using the CMB Stage-IV \cite{S4} 
experiment, planned for observation in early 2020.

\section{Summary} \label{Sec.5}

Based on the recent discovery of the GW event, we have discussed the possibility to test the 
clustering hypothesis of BH binaries similar to the GW150914 event via a network of the 
second-generation GW detectors. Combining with other cosmological probes, we found that with the 3 
year GW observation, the hypothesis of no BH binary clustering will be rejected at more than $3\sigma$ 
significance for a large merger rate, $\dot{n}_0\agt 100$ Gpc$^{-3}$yr$^{-1}$, i.e., $>3\sigma$ 
detection of nonzero signals of the BH binary clustering. For a solid detection of the clustering 
signal, the cross-correlation with galaxies observed by a shallow but wide-field photometric/imaging 
surveys is preferable, and Pan-STARRS would be an ideal survey.

Since the detection significance is almost determined by the cross-correlation with the galaxy 
clustering, the selection bias in the galaxy clustering may affect our results. In photometric galaxy 
measurements, point sources are usually masked. This simultaneously removes the background galaxies at 
the masked regions, and the total signal to noise of the galaxy clustering decreases. However, 
measurement of the galaxy clustering has been well established, and the effect of the selection bias 
can be reduced significantly. Stellar components in the Galaxy contaminate as a false signal which 
could bias the clustering signal, but this effect is negligible in the actual data (e.g., 
Ref.~\cite{DESSV}). 

Once the full operation of the second-generation GW detectors gets started in the coming years, the 
merger rate today, $\dot{n}_0$, will be tightly constrained. Then, measurements of the clustering 
signal of BH binaries and constraints on the clustering bias $b_{\rm BH,0}$ will give us an important 
hint on which type of galaxies the BH binaries are likely to be harbored. 
Since the individual identification of the host galaxy is still challenging with the 
second-generation detectors, the clustering signal of BH binaries would be fruitful and 
complementary information on the formation and evolution of BH binaries. 

In any case, a detection of the spatial clustering of GW sources is an important step toward future 
{\it gravitational-wave cosmology}. As shown in Refs.~\cite{Namikawa:2016,Oguri:2016}, the future 
upgrades of GW detectors such as the ET will be able to precisely measure the clustering of binary 
GW sources at a high statistical significance, from which we can constrain the cosmology, 
complementary to the electromagnetic observations. As one of the representative standard GW sirens, 
BH binaries will also offer a promising cosmological probe, and a measurement of their spatial 
clustering with the second-generation detectors is indispensable for future cosmological study to 
test its feasibility. The prospects of constraining the bias model and cosmology with future GW 
detectors will be investigated in our future work. 

\ifdefined\compile
\begin{acknowledgments}
T.N. is supported by JSPS Postdoctoral Fellowships for Research Abroad No.~26-142. 
A.N. is supported by NSF CAREER Grant No. PHY-1055103 and the H2020-MSCA-RISE- 2015 Grant No. StronGrHEP-690904.
A.T. acknowledges the support from MEXT KAKENHI No.~15H05889. 
\end{acknowledgments}

\bibliographystyle{apsrev}
\bibliography{cite}

\begin{thebibliography}{49}
\expandafter\ifx\csname natexlab\endcsname\relax\def\natexlab#1{#1}\fi
\expandafter\ifx\csname bibnamefont\endcsname\relax
  \def\bibnamefont#1{#1}\fi
\expandafter\ifx\csname bibfnamefont\endcsname\relax
  \def\bibfnamefont#1{#1}\fi
\expandafter\ifx\csname citenamefont\endcsname\relax
  \def\citenamefont#1{#1}\fi
\expandafter\ifx\csname url\endcsname\relax
  \def\url#1{\texttt{#1}}\fi
\expandafter\ifx\csname urlprefix\endcsname\relax\def\urlprefix{URL }\fi
\providecommand{\bibinfo}[2]{#2}
\providecommand{\eprint}[2][]{\url{#2}}

\bibitem[{\citenamefont{{LIGO Scientific and Virgo
  Collaborations}}(2016{\natexlab{a}})}]{GW150914}
\bibinfo{author}{\bibnamefont{{LIGO Scientific and Virgo Collaborations}}},
  \bibinfo{journal}{\prl} \textbf{\bibinfo{volume}{116}},
  \bibinfo{pages}{061102} (\bibinfo{year}{2016}{\natexlab{a}}),
  \href{http://arxiv.org/abs/arXiv:1602.03837}{{\tt arXiv:1602.03837}}.

\bibitem[{\citenamefont{{Evans}}(2014)}]{Evans2014GRG}
\bibinfo{author}{\bibfnamefont{M.}~\bibnamefont{{Evans}}},
  \bibinfo{journal}{Gen. Relativ. Gravit.} \textbf{\bibinfo{volume}{46}},
  \bibinfo{pages}{1778} (\bibinfo{year}{2014}).

\bibitem[{\citenamefont{Punturo et~al.}(2010)}]{Punturo:2010}
\bibinfo{author}{\bibfnamefont{M.}~\bibnamefont{Punturo}} \bibnamefont{et~al.},
  \bibinfo{journal}{Class. Quant. Grav.} \textbf{\bibinfo{volume}{27}},
  \bibinfo{pages}{194002} (\bibinfo{year}{2010}).

\bibitem[{\citenamefont{Dwyer et~al.}(2015)\citenamefont{Dwyer, Sigg, Ballmer,
  Barsotti, Mavalvala, and Evans}}]{Dwyer2015PRD}
\bibinfo{author}{\bibfnamefont{S.}~\bibnamefont{Dwyer}},
  \bibinfo{author}{\bibfnamefont{D.}~\bibnamefont{Sigg}},
  \bibinfo{author}{\bibfnamefont{S.~W.} \bibnamefont{Ballmer}},
  \bibinfo{author}{\bibfnamefont{L.}~\bibnamefont{Barsotti}},
  \bibinfo{author}{\bibfnamefont{N.}~\bibnamefont{Mavalvala}},
  \bibnamefont{and} \bibinfo{author}{\bibfnamefont{M.}~\bibnamefont{Evans}},
  \bibinfo{journal}{\prd} \textbf{\bibinfo{volume}{91}},
  \bibinfo{pages}{082001} (\bibinfo{year}{2015}),
  \href{http://arxiv.org/abs/arXiv:1410.0612}{{\tt arXiv:1410.0612}}.

\bibitem[{\citenamefont{Seoane et~al.}(2013)}]{Seoane:2013qna}
\bibinfo{author}{\bibfnamefont{P.~A.} \bibnamefont{Seoane}}
  \bibnamefont{et~al.} (\bibinfo{collaboration}{eLISA Collaboration})
  (\bibinfo{year}{2013}), \href{http://arxiv.org/abs/1305.5720}{{\tt
  arXiv:1305.5720}}.

\bibitem[{\citenamefont{Sato et~al.}(2009)}]{Sato:2009}
\bibinfo{author}{\bibfnamefont{S.}~\bibnamefont{Sato}} \bibnamefont{et~al.},
  \bibinfo{journal}{Journal of Physics Conference Series}
  \textbf{\bibinfo{volume}{154}}, \bibinfo{pages}{012040}
  (\bibinfo{year}{2009}).

\bibitem[{\citenamefont{Schutz}(1986)}]{Schutz1986}
\bibinfo{author}{\bibfnamefont{B.~F.} \bibnamefont{Schutz}},
  \bibinfo{journal}{\nat} \textbf{\bibinfo{volume}{323}}, \bibinfo{pages}{310}
  (\bibinfo{year}{1986}).

\bibitem[{\citenamefont{Petiteau et~al.}(2011)\citenamefont{Petiteau, Babak,
  and Sesana}}]{Petiteau2011ApJ}
\bibinfo{author}{\bibfnamefont{A.}~\bibnamefont{Petiteau}},
  \bibinfo{author}{\bibfnamefont{S.}~\bibnamefont{Babak}}, \bibnamefont{and}
  \bibinfo{author}{\bibfnamefont{A.}~\bibnamefont{Sesana}},
  \bibinfo{journal}{\apj} \textbf{\bibinfo{volume}{732}}, \bibinfo{pages}{82}
  (\bibinfo{year}{2011}), \href{http://arxiv.org/abs/arXiv:1102.0769}{{\tt
  arXiv:1102.0769}}.

\bibitem[{\citenamefont{Cutler and Holz}(2009)}]{Cutler:2009qv}
\bibinfo{author}{\bibfnamefont{C.}~\bibnamefont{Cutler}} \bibnamefont{and}
  \bibinfo{author}{\bibfnamefont{D.~E.} \bibnamefont{Holz}},
  \bibinfo{journal}{\prd} \textbf{\bibinfo{volume}{80}},
  \bibinfo{pages}{104009} (\bibinfo{year}{2009}),
  \href{http://arxiv.org/abs/arXiv:0906.3752}{{\tt arXiv:0906.3752}}.

\bibitem[{\citenamefont{Nishizawa et~al.}(2012)\citenamefont{Nishizawa, Yagi,
  Taruya, and Tanaka}}]{Nishizawa:2011eq}
\bibinfo{author}{\bibfnamefont{A.}~\bibnamefont{Nishizawa}},
  \bibinfo{author}{\bibfnamefont{K.}~\bibnamefont{Yagi}},
  \bibinfo{author}{\bibfnamefont{A.}~\bibnamefont{Taruya}}, \bibnamefont{and}
  \bibinfo{author}{\bibfnamefont{T.}~\bibnamefont{Tanaka}},
  \bibinfo{journal}{\prd} \textbf{\bibinfo{volume}{85}},
  \bibinfo{pages}{044047} (\bibinfo{year}{2012}),
  \href{http://arxiv.org/abs/arXiv:1110.2865}{{\tt arXiv:1110.2865}}.

\bibitem[{\citenamefont{Sathyaprakash et~al.}(2010)\citenamefont{Sathyaprakash,
  Schutz, and Van Den~Broeck}}]{Sathyaprakash:2009xt}
\bibinfo{author}{\bibfnamefont{B.~S.} \bibnamefont{Sathyaprakash}},
  \bibinfo{author}{\bibfnamefont{B.~F.} \bibnamefont{Schutz}},
  \bibnamefont{and} \bibinfo{author}{\bibfnamefont{C.}~\bibnamefont{Van
  Den~Broeck}}, \bibinfo{journal}{Class. Quant. Grav.}
  \textbf{\bibinfo{volume}{27}}, \bibinfo{pages}{215006}
  (\bibinfo{year}{2010}), \href{http://arxiv.org/abs/arXiv:0906.4151}{{\tt
  arXiv:0906.4151}}.

\bibitem[{\citenamefont{Taylor and Gair}(2012)}]{Taylor:2012db}
\bibinfo{author}{\bibfnamefont{S.~R.} \bibnamefont{Taylor}} \bibnamefont{and}
  \bibinfo{author}{\bibfnamefont{J.~R.} \bibnamefont{Gair}},
  \bibinfo{journal}{\prd} \textbf{\bibinfo{volume}{86}},
  \bibinfo{pages}{023502} (\bibinfo{year}{2012}),
  \href{http://arxiv.org/abs/arXiv:1204.6739}{{\tt arXiv:1204.6739}}.

\bibitem[{\citenamefont{Hirata et~al.}(2010)\citenamefont{Hirata, Holz, and
  Cutler}}]{Hirata:2010}
\bibinfo{author}{\bibfnamefont{C.~M.} \bibnamefont{Hirata}},
  \bibinfo{author}{\bibfnamefont{D.~E.} \bibnamefont{Holz}}, \bibnamefont{and}
  \bibinfo{author}{\bibfnamefont{C.}~\bibnamefont{Cutler}},
  \bibinfo{journal}{\prd} \textbf{\bibinfo{volume}{81}},
  \bibinfo{pages}{124046} (\bibinfo{year}{2010}),
  \href{http://arxiv.org/abs/arXiv:1004.3988}{{\tt arXiv:1004.3988}}.

\bibitem[{\citenamefont{Shang and Haiman}(2011)}]{Shang2011MNRAS}
\bibinfo{author}{\bibfnamefont{C.}~\bibnamefont{Shang}} \bibnamefont{and}
  \bibinfo{author}{\bibfnamefont{Z.}~\bibnamefont{Haiman}},
  \bibinfo{journal}{\mnras} \textbf{\bibinfo{volume}{411}}, \bibinfo{pages}{9}
  (\bibinfo{year}{2011}), \href{http://arxiv.org/abs/arXiv:1004.3562}{{\tt
  arXiv:1004.3562}}.

\bibitem[{\citenamefont{Camera and Nishizawa}(2013)}]{Camera:2013xfa}
\bibinfo{author}{\bibfnamefont{S.}~\bibnamefont{Camera}} \bibnamefont{and}
  \bibinfo{author}{\bibfnamefont{A.}~\bibnamefont{Nishizawa}},
  \bibinfo{journal}{\prl} \textbf{\bibinfo{volume}{110}},
  \bibinfo{pages}{151103} (\bibinfo{year}{2013}),
  \href{http://arxiv.org/abs/arXiv:1303.5446}{{\tt arXiv:1303.5446}}.

\bibitem[{\citenamefont{Namikawa et~al.}(2016)\citenamefont{Namikawa,
  Nishizawa, and Taruya}}]{Namikawa:2016}
\bibinfo{author}{\bibfnamefont{T.}~\bibnamefont{Namikawa}},
  \bibinfo{author}{\bibfnamefont{A.}~\bibnamefont{Nishizawa}},
  \bibnamefont{and} \bibinfo{author}{\bibfnamefont{A.}~\bibnamefont{Taruya}},
  \bibinfo{journal}{\prl} \textbf{\bibinfo{volume}{116}},
  \bibinfo{pages}{121302} (\bibinfo{year}{2016}),
  \href{http://arxiv.org/abs/arXiv:1511.04638}{{\tt arXiv:1511.04638}}.

\bibitem[{\citenamefont{Oguri}(2016)}]{Oguri:2016}
\bibinfo{author}{\bibfnamefont{M.}~\bibnamefont{Oguri}},
  \bibinfo{journal}{\prd} \textbf{\bibinfo{volume}{93}},
  \bibinfo{pages}{083511} (\bibinfo{year}{2016}),
  \href{http://arxiv.org/abs/arXiv:1603.02356}{{\tt arXiv:1603.02356}}.

\bibitem[{\citenamefont{Clesse and Garcia-Bellido}(2015)}]{Clesse:2015}
\bibinfo{author}{\bibfnamefont{S.}~\bibnamefont{Clesse}} \bibnamefont{and}
  \bibinfo{author}{\bibfnamefont{J.}~\bibnamefont{Garcia-Bellido}},
  \bibinfo{journal}{\prd} \textbf{\bibinfo{volume}{92}},
  \bibinfo{pages}{023524} (\bibinfo{year}{2015}),
  \href{http://arxiv.org/abs/arXiv:1501.07565}{{\tt arXiv:1501.07565}}.

\bibitem[{\citenamefont{Bird et~al.}(2016)\citenamefont{Bird, Cholis, Munoz,
  Ali-Haimoud, Kamionkowski et~al.}}]{Bird:2016}
\bibinfo{author}{\bibfnamefont{S.}~\bibnamefont{Bird}},
  \bibinfo{author}{\bibfnamefont{I.}~\bibnamefont{Cholis}},
  \bibinfo{author}{\bibfnamefont{J.~B.} \bibnamefont{Munoz}},
  \bibinfo{author}{\bibfnamefont{Y.}~\bibnamefont{Ali-Haimoud}},
  \bibinfo{author}{\bibfnamefont{M.}~\bibnamefont{Kamionkowski}},
  \bibnamefont{et~al.}, \bibinfo{journal}{\prl} \textbf{\bibinfo{volume}{116}},
  \bibinfo{pages}{201301} (\bibinfo{year}{2016}),
  \href{http://arxiv.org/abs/arXiv:1603.00464}{{\tt arXiv:1603.00464}}.

\bibitem[{\citenamefont{Clesse and Garcia-Bellido}(2016)}]{Clesse:2016}
\bibinfo{author}{\bibfnamefont{S.}~\bibnamefont{Clesse}} \bibnamefont{and}
  \bibinfo{author}{\bibfnamefont{J.}~\bibnamefont{Garcia-Bellido}}
  (\bibinfo{year}{2016}), \href{http://arxiv.org/abs/arXiv:1603.05235}{{\tt
  arXiv:1603.05235}}.

\bibitem[{\citenamefont{Belczynski et~al.}(2010)\citenamefont{Belczynski,
  Dominik, Bulik, O'Shaughnessy, Fryer, and Holz}}]{Bekczynski:2010}
\bibinfo{author}{\bibfnamefont{K.}~\bibnamefont{Belczynski}},
  \bibinfo{author}{\bibfnamefont{M.}~\bibnamefont{Dominik}},
  \bibinfo{author}{\bibfnamefont{T.}~\bibnamefont{Bulik}},
  \bibinfo{author}{\bibfnamefont{R.}~\bibnamefont{O'Shaughnessy}},
  \bibinfo{author}{\bibfnamefont{C.}~\bibnamefont{Fryer}}, \bibnamefont{and}
  \bibinfo{author}{\bibfnamefont{D.~E.} \bibnamefont{Holz}},
  \bibinfo{journal}{\apj} \textbf{\bibinfo{volume}{715}}, \bibinfo{pages}{L138}
  (\bibinfo{year}{2010}), \href{http://arxiv.org/abs/arXiv:1004.0386}{{\tt
  arXiv:1004.0386}}.

\bibitem[{\citenamefont{Kinugawa et~al.}(2014)\citenamefont{Kinugawa, Inayoshi,
  Hotokezaka, Nakauchi, and Nakamura}}]{Kinugawa:2014}
\bibinfo{author}{\bibfnamefont{T.}~\bibnamefont{Kinugawa}},
  \bibinfo{author}{\bibfnamefont{K.}~\bibnamefont{Inayoshi}},
  \bibinfo{author}{\bibfnamefont{K.}~\bibnamefont{Hotokezaka}},
  \bibinfo{author}{\bibfnamefont{D.}~\bibnamefont{Nakauchi}}, \bibnamefont{and}
  \bibinfo{author}{\bibfnamefont{T.}~\bibnamefont{Nakamura}},
  \bibinfo{journal}{\mnras} \textbf{\bibinfo{volume}{442}},
  \bibinfo{pages}{2963} (\bibinfo{year}{2014}),
  \href{http://arxiv.org/abs/arXiv:1402.6672}{{\tt arXiv:1402.6672}}.

\bibitem[{\citenamefont{Lewis et~al.}(2000)\citenamefont{Lewis, Challinor, and
  Lasenby}}]{CAMB}
\bibinfo{author}{\bibfnamefont{A.}~\bibnamefont{Lewis}},
  \bibinfo{author}{\bibfnamefont{A.}~\bibnamefont{Challinor}},
  \bibnamefont{and} \bibinfo{author}{\bibfnamefont{A.}~\bibnamefont{Lasenby}},
  \bibinfo{journal}{\apj} \textbf{\bibinfo{volume}{538}}, \bibinfo{pages}{473}
  (\bibinfo{year}{2000}),
  \href{http://arxiv.org/abs/arXiv:astro-ph/9911177}{{\tt
  arXiv:astro-ph/9911177}}.

\bibitem[{\citenamefont{Komatsu et~al.}(2011)}]{Komatsu:2010fb}
\bibinfo{author}{\bibfnamefont{E.}~\bibnamefont{Komatsu}} \bibnamefont{et~al.},
  \bibinfo{journal}{\apj} \textbf{\bibinfo{volume}{192}}, \bibinfo{pages}{18}
  (\bibinfo{year}{2011}), \href{http://arxiv.org/abs/arXiv:1001.4538}{{\tt
  arXiv:1001.4538}}.

\bibitem[{\citenamefont{Smith et~al.}(2003)}]{Smith:2003}
\bibinfo{author}{\bibfnamefont{R.~E.} \bibnamefont{Smith}}
  \bibnamefont{et~al.}, \bibinfo{journal}{\mnras}
  \textbf{\bibinfo{volume}{341}}, \bibinfo{pages}{1311} (\bibinfo{year}{2003}),
  \href{http://arxiv.org/abs/arXiv:astro-ph/0207664}{{\tt
  arXiv:astro-ph/0207664}}.

\bibitem[{\citenamefont{Takahashi et~al.}(2012)}]{Takahashi:2012}
\bibinfo{author}{\bibfnamefont{R.}~\bibnamefont{Takahashi}}
  \bibnamefont{et~al.}, \bibinfo{journal}{\apj} \textbf{\bibinfo{volume}{761}},
  \bibinfo{pages}{152} (\bibinfo{year}{2012}),
  \href{http://arxiv.org/abs/arXiv:1208.2701}{{\tt arXiv:1208.2701}}.

\bibitem[{\citenamefont{Hu}(2001)}]{Hu:2001fb}
\bibinfo{author}{\bibfnamefont{W.}~\bibnamefont{Hu}}, \bibinfo{journal}{\prd}
  \textbf{\bibinfo{volume}{65}}, \bibinfo{pages}{023003}
  (\bibinfo{year}{2001}), \href{http://arxiv.org/abs/astro-ph/0108090}{{\tt
  astro-ph/0108090}}.

\bibitem[{\citenamefont{Namikawa et~al.}(2011)\citenamefont{Namikawa, Okamura,
  and Taruya}}]{Namikawa:2011}
\bibinfo{author}{\bibfnamefont{T.}~\bibnamefont{Namikawa}},
  \bibinfo{author}{\bibfnamefont{T.}~\bibnamefont{Okamura}}, \bibnamefont{and}
  \bibinfo{author}{\bibfnamefont{A.}~\bibnamefont{Taruya}},
  \bibinfo{journal}{\prd} \textbf{\bibinfo{volume}{83}},
  \bibinfo{pages}{123514} (\bibinfo{year}{2011}),
  \href{http://arxiv.org/abs/arXiv:1103.1118}{{\tt arXiv:1103.1118}}.

\bibitem[{\citenamefont{Yamauchi et~al.}(2013)\citenamefont{Yamauchi, Namikawa,
  and Taruya}}]{Yamauchi:2013}
\bibinfo{author}{\bibfnamefont{D.}~\bibnamefont{Yamauchi}},
  \bibinfo{author}{\bibfnamefont{T.}~\bibnamefont{Namikawa}}, \bibnamefont{and}
  \bibinfo{author}{\bibfnamefont{A.}~\bibnamefont{Taruya}},
  \bibinfo{journal}{\jcap} \textbf{\bibinfo{volume}{08}}, \bibinfo{pages}{051}
  (\bibinfo{year}{2013}), \href{http://arxiv.org/abs/arXiv:1305.3348}{{\tt
  arXiv:1305.3348}}.

\bibitem[{\citenamefont{{LIGO Scientific and Virgo
  Collaborations}}(2016{\natexlab{b}})}]{GW150914rate}
\bibinfo{author}{\bibnamefont{{LIGO Scientific and Virgo Collaborations}}}
  (\bibinfo{year}{2016}{\natexlab{b}}),
  \href{http://arxiv.org/abs/arXiv:1602.03842}{{\tt arXiv:1602.03842}}.

\bibitem[{\citenamefont{Rassat et~al.}(2008)}]{Rassat:2008}
\bibinfo{author}{\bibfnamefont{A.}~\bibnamefont{Rassat}} \bibnamefont{et~al.}
  (\bibinfo{year}{2008}), \href{http://arxiv.org/abs/arXiv:0810.0003}{{\tt
  arXiv:0810.0003}}.

\bibitem[{\citenamefont{{Euclid Theory Working Group}}(2013)}]{Euclid}
\bibinfo{author}{\bibnamefont{{Euclid Theory Working Group}}},
  \bibinfo{journal}{Living Rev. Relativity} \textbf{\bibinfo{volume}{16}},
  \bibinfo{pages}{6} (\bibinfo{year}{2013}),
  \href{http://arxiv.org/abs/arXiv:1206.1225}{{\tt arXiv:1206.1225}}.

\bibitem[{\citenamefont{Bartelmann and Schneider}(2001)}]{Bartelmann:2001}
\bibinfo{author}{\bibfnamefont{M.}~\bibnamefont{Bartelmann}} \bibnamefont{and}
  \bibinfo{author}{\bibfnamefont{P.}~\bibnamefont{Schneider}},
  \bibinfo{journal}{Phys. Rep.} \textbf{\bibinfo{volume}{340}},
  \bibinfo{pages}{291} (\bibinfo{year}{2001}),
  \href{http://arxiv.org/abs/arXiv:astro-ph/9912508}{{\tt
  arXiv:astro-ph/9912508}}.

\bibitem[{\citenamefont{Lewis and Challinor}(2006)}]{Lewis:2006fu}
\bibinfo{author}{\bibfnamefont{A.}~\bibnamefont{Lewis}} \bibnamefont{and}
  \bibinfo{author}{\bibfnamefont{A.}~\bibnamefont{Challinor}},
  \bibinfo{journal}{Phys. Rep.} \textbf{\bibinfo{volume}{429}},
  \bibinfo{pages}{1} (\bibinfo{year}{2006}),
  \href{http://arxiv.org/abs/arXiv:astro-ph/0601594}{{\tt
  arXiv:astro-ph/0601594}}.

\bibitem[{\citenamefont{Hu and Okamoto}(2002)}]{Hu:2001kj}
\bibinfo{author}{\bibfnamefont{W.}~\bibnamefont{Hu}} \bibnamefont{and}
  \bibinfo{author}{\bibfnamefont{T.}~\bibnamefont{Okamoto}},
  \bibinfo{journal}{\apj} \textbf{\bibinfo{volume}{574}}, \bibinfo{pages}{566}
  (\bibinfo{year}{2002}), \href{http://arxiv.org/abs/astro-ph/0111606}{{\tt
  astro-ph/0111606}}.

\bibitem[{\citenamefont{Chisari et~al.}(2014)\citenamefont{Chisari, Dvorkin,
  and Schmidt}}]{Chisari:2014}
\bibinfo{author}{\bibfnamefont{N.~E.} \bibnamefont{Chisari}},
  \bibinfo{author}{\bibfnamefont{C.}~\bibnamefont{Dvorkin}}, \bibnamefont{and}
  \bibinfo{author}{\bibfnamefont{F.}~\bibnamefont{Schmidt}},
  \bibinfo{journal}{\prd} \textbf{\bibinfo{volume}{90}},
  \bibinfo{pages}{043527} (\bibinfo{year}{2014}),
  \href{http://arxiv.org/abs/arXiv:1406.4871}{{\tt arXiv:1406.4871}}.

\bibitem[{\citenamefont{{\textsc{Polarbear} Collaboration}}(2014)}]{PB14:phi}
\bibinfo{author}{\bibnamefont{{\textsc{Polarbear} Collaboration}}},
  \bibinfo{journal}{\prl} \textbf{\bibinfo{volume}{113}},
  \bibinfo{pages}{021301} (\bibinfo{year}{2014}).

\bibitem[{\citenamefont{Knox}(1999)}]{Knox:1999}
\bibinfo{author}{\bibfnamefont{L.}~\bibnamefont{Knox}}, \bibinfo{journal}{\prd}
  \textbf{\bibinfo{volume}{60}}, \bibinfo{pages}{103516}
  (\bibinfo{year}{1999}),
  \href{http://arxiv.org/abs/arXiv:astro-ph/9902046}{{\tt
  arXiv:astro-ph/9902046}}.

\bibitem[{\citenamefont{Sidery et~al.}(2014)\citenamefont{Sidery, Aylott,
  Christensen, Farr, Farr, Feroz, Gair, Grover, Graff, Hanna
  et~al.}}]{Sidery2014PRD}
\bibinfo{author}{\bibfnamefont{T.}~\bibnamefont{Sidery}},
  \bibinfo{author}{\bibfnamefont{B.}~\bibnamefont{Aylott}},
  \bibinfo{author}{\bibfnamefont{N.}~\bibnamefont{Christensen}},
  \bibinfo{author}{\bibfnamefont{B.}~\bibnamefont{Farr}},
  \bibinfo{author}{\bibfnamefont{W.}~\bibnamefont{Farr}},
  \bibinfo{author}{\bibfnamefont{F.}~\bibnamefont{Feroz}},
  \bibinfo{author}{\bibfnamefont{J.}~\bibnamefont{Gair}},
  \bibinfo{author}{\bibfnamefont{K.}~\bibnamefont{Grover}},
  \bibinfo{author}{\bibfnamefont{P.}~\bibnamefont{Graff}},
  \bibinfo{author}{\bibfnamefont{C.}~\bibnamefont{Hanna}},
  \bibnamefont{et~al.}, \bibinfo{journal}{\prd} \textbf{\bibinfo{volume}{89}},
  \bibinfo{pages}{084060} (\bibinfo{year}{2014}),
  \href{http://arxiv.org/abs/arXiv:1312.6013}{{\tt arXiv:1312.6013}}.

\bibitem[{\citenamefont{{Dark Energy Survey Collaboration}}(2006)}]{DES}
\bibinfo{author}{\bibnamefont{{Dark Energy Survey Collaboration}}}
  (\bibinfo{year}{2006}), \urlprefix\url{http://www.darkenergysurvey.org/}.

\bibitem[{\citenamefont{Smith et~al.}(2012)}]{Smith:2010gu}
\bibinfo{author}{\bibfnamefont{K.~M.} \bibnamefont{Smith}}
  \bibnamefont{et~al.}, \bibinfo{journal}{\jcap} \textbf{\bibinfo{volume}{06}},
  \bibinfo{pages}{014} (\bibinfo{year}{2012}),
  \href{http://arxiv.org/abs/arXiv:1010.0048}{{\tt arXiv:1010.0048}}.

\bibitem[{\citenamefont{Hirata and Seljak}(2003)}]{Hirata:2003ka}
\bibinfo{author}{\bibfnamefont{C.~M.} \bibnamefont{Hirata}} \bibnamefont{and}
  \bibinfo{author}{\bibfnamefont{U.}~\bibnamefont{Seljak}},
  \bibinfo{journal}{\prd} \textbf{\bibinfo{volume}{68}},
  \bibinfo{pages}{083002} (\bibinfo{year}{2003}),
  \href{http://arxiv.org/abs/arXiv:0306354}{{\tt arXiv:0306354}}.

\bibitem[{\citenamefont{Cai et~al.}(2009)\citenamefont{Cai, Angulo, Baugh,
  Cole, Frenk, and Jenkins}}]{Cai:2009}
\bibinfo{author}{\bibfnamefont{Y.-C.} \bibnamefont{Cai}},
  \bibinfo{author}{\bibfnamefont{R.~E.} \bibnamefont{Angulo}},
  \bibinfo{author}{\bibfnamefont{C.~M.} \bibnamefont{Baugh}},
  \bibinfo{author}{\bibfnamefont{S.}~\bibnamefont{Cole}},
  \bibinfo{author}{\bibfnamefont{C.~S.} \bibnamefont{Frenk}}, \bibnamefont{and}
  \bibinfo{author}{\bibfnamefont{A.}~\bibnamefont{Jenkins}},
  \bibinfo{journal}{\mnras} \textbf{\bibinfo{volume}{395}},
  \bibinfo{pages}{1185} (\bibinfo{year}{2009}),
  \href{http://arxiv.org/abs/arXiv:0810.2300}{{\tt arXiv:0810.2300}}.

\bibitem[{\citenamefont{Sathyaprakash and Schutz}(2009)}]{Sathyaprakash:2009}
\bibinfo{author}{\bibfnamefont{B.~S.} \bibnamefont{Sathyaprakash}}
  \bibnamefont{and} \bibinfo{author}{\bibfnamefont{B.~F.}
  \bibnamefont{Schutz}}, \bibinfo{journal}{Living Rev. Relat.}
  \textbf{\bibinfo{volume}{12}}, \bibinfo{pages}{2} (\bibinfo{year}{2009}),
  \href{http://arxiv.org/abs/arXiv:0903.0338}{{\tt arXiv:0903.0338}}.

\bibitem[{\citenamefont{Calabrese et~al.}(2014)}]{AdvACT}
\bibinfo{author}{\bibfnamefont{E.}~\bibnamefont{Calabrese}}
  \bibnamefont{et~al.}, \bibinfo{journal}{JCAP} \textbf{\bibinfo{volume}{08}},
  \bibinfo{pages}{010} (\bibinfo{year}{2014}),
  \href{http://arxiv.org/abs/arXiv:1406.4794}{{\tt arXiv:1406.4794}}.

\bibitem[{\citenamefont{Suzuki et~al.}(2016)}]{SA}
\bibinfo{author}{\bibfnamefont{A.}~\bibnamefont{Suzuki}} \bibnamefont{et~al.}
  (\bibinfo{collaboration}{\textsc{POLARBEAR} Collaboration}),
  \bibinfo{journal}{J. Low Temp. Phys.} \textbf{\bibinfo{volume}{184}},
  \bibinfo{pages}{805} (\bibinfo{year}{2016}),
  \href{http://arxiv.org/abs/arXiv:1512.07299}{{\tt arXiv:1512.07299}}.

\bibitem[{\citenamefont{{LSST Dark Energy Science Collaboration}}(2012)}]{LSST}
\bibinfo{author}{\bibnamefont{{LSST Dark Energy Science Collaboration}}}
  (\bibinfo{year}{2012}), \href{http://arxiv.org/abs/arXiv:1211.0310}{{\tt
  arXiv:1211.0310}}.

\bibitem[{\citenamefont{Abazajian et~al.}(2015)}]{S4}
\bibinfo{author}{\bibfnamefont{K.}~\bibnamefont{Abazajian}}
  \bibnamefont{et~al.}, \bibinfo{journal}{Astropart. Phys.}
  \textbf{\bibinfo{volume}{63}}, \bibinfo{pages}{66} (\bibinfo{year}{2015}),
  \href{http://arxiv.org/abs/arXiv:1309.5383}{{\tt arXiv:1309.5383}}.

\bibitem[{\citenamefont{Crocce et~al.}(2015)}]{DESSV}
\bibinfo{author}{\bibfnamefont{M.}~\bibnamefont{Crocce}} \bibnamefont{et~al.},
  \bibinfo{journal}{\mnras} \textbf{\bibinfo{volume}{455}},
  \bibinfo{pages}{4301} (\bibinfo{year}{2015}).

\end{thebibliography}
\fi

\end{document}